\documentclass[aps,pre,superscriptaddress,showpacs,amsmath,amssymb,twocolumn]{revtex4}

\usepackage{graphicx}
\usepackage{bm}% bold math
\usepackage{color}
\usepackage{amsmath}

\def\Vect#1{\mbox{\boldmath $#1$}}
\def\cl{{\cal L}}
\def\dgam{\dot{\gamma}(t)}

\def\bs#1{\boldsymbol{#1}}
\begin{document}

\title{Nonequilibrium identities and response theory for dissipative particles}

\author{Hisao Hayakawa }
\email[]{hisao@yukawa.kyoto-u.ac.jp}
\affiliation{Yukawa Institute for Theoretical Physics, Kyoto University, Kyoto 606-8502, Japan}
\author{Michio Otsuki}
\affiliation{Department of Materials Science, Shimane University, Matsue 690-8504, Japan}

%\publishedin{%         %Write this ONLY in cases of addenda and errata
%Prog.~Theor.~Phys.\ \textbf{XX} (19YY), page.}

%\recdate{Mmmmm DD, YYYY}%            %Editorial Office will fill in this.

\begin{abstract}
%{%         %this abstract is neglected when [addenda] or [errata]
We derive some nonequilibrium identities such as the integral fluctuation theorem and the Jarzynski equality starting from a nonequilibrium state for dissipative classical systems.
Thanks to the existence of the integral fluctuation theorem we can naturally introduce an entropy-like quantity for dissipative classical systems in far from equilibrium states.  
 We also derive  the generalized Green-Kubo formula as a nonlinear response theory for a steady dynamics around a nonequilibrium  state. 
We numerically verify  the validity of the derived formulas for sheared frictionless granular particles.

%We verify the validity of the derived formulae in terms of the molecular dynamics simulation.

\end{abstract}
%\date{\today}


\maketitle

\section{Introduction}

Construction of a nonlinear response theory around a nonequilibrium state is one of the most challenging problems in theoretical physics.~\cite{Zubarev74,McLennan88,Evans08}.
The most remarkable achievements in the last two decades are the generalized Green-Kubo relation~\cite{Morriss87}
and various fluctuation theorems~\cite{FT93,GC95,Jarzynski,Crooks,Kurchan,Hatano-Sasa,Evans02,Seifert12} even in quantum systems~\cite{Esposito09}.
These relations can reproduce Green-Kubo formula and the reciprocal relation in the linearized limit, and give a mechanical foundation of the second law of the thermodynamics.
On the other hand, there are some cases that the heat fluctuation theorem is, at least, no longer valid for some situations \cite{Ren10,kanazawa13}.
Thus, it is important to know whether these identities are valid in arbitrary nonequilibrium situations.

Although it has been believed 
that these identities such as the fluctuation theorem are supported by the local time-reversal symmetry or the detailed balance condition,
some experiments suggest the existence of the fluctuation theorem or related identities even in granular systems which do not have neither any time reversal symmetry nor detailed balance~\cite{mennon04,kumar11,joubaud12,Naert12,mounier12}, though there exists a counter argument \cite{Puglisi05}.
It is remarkable that series of papers by Puglisi and his coworkers \cite{Puglisi05EPL,Puglisi06PRE,Puglisi06JSM,Sarracino10} clarified that granular fluids do not hold the conventional fluctuation theorem but have only the second type fluctuation theorem by Evans and Searles~\cite{Evans94}.  
We note that Chong et al. have proven the existence of 
both the generalized Green-Kubo relation and the integral fluctuation theorem \cite{Seifert05,Seifert12} for a granular system under a steady plane shear~\cite{Chong09b}.
They also developed the representation of a nonequilibrium steady-state distribution function~\cite{Chong10} and the liquid theory for sheared dense granular systems \cite{Hayakawa10a}.
%Recently, one of the present authors~\cite{Hayakawa13} extended their previous formulation to the nonlinear response theory around a nonequilibrium steady state for vibrating granular beds. 

Unfortunately, the achievement of the previous studies~\cite{Chong09b,Chong10,Hayakawa10a} for sheared granular systems is not a response theory from a nonequilibrium state but a relaxation theory to a nonequilibrium steady state starting from an equilibrium state.
One may regard the previous results as the relations only valid for sheared granular fluids, but our formulation is applicable to any classical dissipative systems.
Thus, one of the main purposes of this paper is to extend the previous formulation to obtain the general nonlinear response theory from any nonequilibrium state for dissipative classical particles, in which the system may not have neither any local time reversal symmetry nor local detailed balance.
%However, we should note that our formulation can be used for any dissipative classical systems which include dissipations and external forces.
We also demonstrate the validity of the derived formulas from the direct numerical simulation of sheared frictionless granular particles.

The organization of this paper is as follows.
In Sec. II, we summarize the general framework of our analysis in terms of  Liouville equation for a system of dissipative classical particles and introduce some known identities which will be used in this paper.
%, and present some results for a simpler situation which can be described by the combination of two steady Liouville operators (subsection II B).
Section III which is the main part of this paper consists of four parts. 
In the first part (Sec. III A) we derive the integral fluctuation theorem (IFT).  
We also demonstrate the existence of a positive definite quantity during the time evolution, which can be regarded as the entropy production rate in the system.
In the second part (Sec. III B) we discuss how to derive the conventional fluctuation theorem for the rate of a dissipation function without local time reversal symmetry.
In the third part (Sec. III C) we derive the Jarzynski equality if the evolution dynamics contains a part not described by the Liouvillian.
In the fourth part (Sec. III D), we derive the generalized Green-Kubo formula for a steady dynamics.
In Sec. IV we simulate sheared frictionless granular particles to verify the identities derived in Sec. III.
In Sec. V we discuss our results and explain the physical implication of our results toward the construction of a statistical mechanics for a classical dissipative system in a nonequilibrium steady state. 
Finally we summarize our conclusion in Sec. VI.
In Appendix A, we introduce an alternative derivation of IFT.
In Appendix B, we explain the derivation of the Jarzynski equality when we start from the canonical distribution to clarify its physical meaning.
In Appendix C, we discuss the effect of the measurement to get the generalized Jarzynski equality.

%\section{Some identities Liouville equation}

\section{General framework}

Let us consider a system of $N$ identical soft spherical particles characterized by the positions $\{ \bs{r}_i(t)\}_{i=1}^N$ and the momenta  $\{\bs{p}_i(t)\}_{i=1}^N$ of particles at time $t$. 
We assume that the system involves dissipations and external forces to relax a steady state.
For a while, we do not specify the system we consider, and we discuss universal results which can be used for any dissipative classical systems in this and the next sections.

The basic equation for the statistical mechanics of such particles is  the Liouville equation \cite{Evans08,brey97,dufty06,dufty06b,HO2008,Chong_etal_in_preparation,Suzuki-Hayakawa,vanNojie}.
%The Liouville equation for granular fluids was investigated by Brey {\it et al.}\cite{brey97,dufty06,dufty06b} some time ago.
The argument in this section is parallel to that in Ref. \cite{Evans08}.
Let $i\cl (t)$ be the total Liouvillian which operates  a physical function or an observable $A(\Vect{\Gamma}(t))$ 
starting from $t=0$ as
\begin{equation}\label{Liouville}
\frac{dA(\Vect{\Gamma}(t))}{d{t}}=U_\rightarrow(0,t) i \cl(t) A(\Vect{\Gamma}) , 
{~~} 
A(\Vect{\Gamma}(t))=U_\rightarrow(0,t)
%T_\rightarrow e^{i\int_{-\infty}^td\tau \cl_{\rm tot}(\tau)}
 A(\Vect{\Gamma}), 
\end{equation}
where 
%\begin{eqnarray}
$
U_\rightarrow(0,t) \equiv T_\rightarrow e^{i \int_{0}^tds \cl(s)}
$ 
is the time evolution operator
%=
%\sum_{n=0}^{\infty}
%\int_{0}^t ds_1 \int_{0}^{s_1}ds_2 
%\cdots \int_{0}^{s_{n-1}} ds_n 
%i{\cal L}(s_n) \cdots i {\cal L}(s_2)i {\cal L}(s_1)
%,$
with the time ordered exponential function 
$T_\rightarrow e^{i \int_{0}^tds \cl(s)}
=
\sum_{n=0}^{\infty}
\int_{0}^t ds_1 \int_{0}^{s_1}ds_2 
\cdots \int_{0}^{s_{n-1}} ds_n 
i{\cal L}(s_n) \cdots i {\cal L}(s_2)i {\cal L}(s_1)
,$
 and
$\Vect{\Gamma}(t)=\{\boldsymbol{r}_i(t),\boldsymbol{p}_i(t)\}_{i=1}^N$ 
represents all phase variables
%is the set of the positions $\boldsymbol{r}_i(t)$ and the momenta $\boldsymbol{p}_i(t)$ for $N$ particles 
with the abbreviation $\Vect{\Gamma}\equiv \Vect{\Gamma}(0)$.  
We note that there are some trivial relations for $U_\rightarrow(t_0,t)$ such as
%\begin{equation}\label{eq:2.2}
$U_\rightarrow(t_0,t)=U_\rightarrow(t_0,s)U_\rightarrow(s,t)$ for $t_0\le s\le t$ and  
$U_\rightarrow(t_0,t) \tilde{ f}(\Vect{\Gamma}(t_0))=\tilde{ f}(\Vect{\Gamma}(t))$ 
%\end{equation}
for an arbitrary function $\tilde{ f}(\Vect{\Gamma}(t))$.

The total Liouvillian consists of three parts, the elastic part, the dissipative or the viscous part and the part from an external force. 
We write $i\cl(t)$ as 
\begin{eqnarray}
i\cl(t)&=&
\dot{\bs{\Gamma}}(\bs{\Gamma},t)\cdot\frac{\partial}{\partial \bs{\Gamma}} \nonumber\\
&=&i\cl^{\rm (el)}(\bs{\Gamma})+i\cl^{\rm (vis)}(\bs{\Gamma})+i\cl_{\rm ext}(\bs{\Gamma},t),
\end{eqnarray}
 where
$i\cl^{\rm (el)}$ is  the sum of the free part and the elastic collision part, 
\begin{equation}\label{cl}
i\cl^{\rm (el)}(\bs{\Gamma})=\sum_{i=1}^N\frac{\bs{p}_i}{m}\cdot\frac{\partial}{\partial \boldsymbol{r}_i}+\bs{F}^{\rm (el)}_i\cdot\frac{\partial}{\partial \bs{p}_i} 
\end{equation}
with the mass $m$ of each particle and the elastic or the conservative force $\bs{F}^{(\rm el)}_i$ acting on the $i-$th particle which is assumed to be represented by the sum of the pairwise elastic force $\bs{F}^{\rm (el)}_{ij}$ as 
$\bs{F}_{i}^{\rm (el)}=\sum_{j\ne i}\bs{F}^{\rm (el)}_{ij}$.
Here, the pairwise elastic force can be represented by the potential $V(r_{ij})$ as $\bs{F}^{\rm (el)}_{ij}=-\partial V(r_{ij})/\partial \bs{r}_{ij}$, 
where $\bs{r}_{ij}\equiv \bs{r}_i-\bs{r}_j$ and $r_{ij}\equiv |\bs{r}_{ij}|$.
%with the peculiar momentum $\boldsymbol{p}_i$ of the $i-$th particle and the elastic force $\bs{F}_i^{\rm (el)}=\sum_{j\ne i} \bs{F}_{ij}^{\rm (el)}$ acting on $i$-th particle with the force $\bs{F}_{ij}^{\rm (el)}$ acting from $j$-th particle to $i$-th particle. 
Similarly, the viscous Liouvillian $i\cl^{\rm (vis)}(\bs{\Gamma})$ is given by
\begin{equation}\label{viscous-L}
i\cl^{\rm (vis)}(\bs{\Gamma})=\sum_{i=1}^N \bs{F}^{\rm (vis)}_i(\bs{\Gamma})\cdot\frac{\partial}{\partial \bs{p}_i},
\end{equation}
where $\bs{F}^{\rm (vis)}_i$ is the viscous force acting on $i-$th particle.
For simplicity, we assume that $\bs{F}^{\rm (vis)}_i(\bs{\Gamma})$ does not depend on the time explicitly, which can be a function of time through the change of $\bs{\Gamma}(t)$.
The explicit form of the Liouvillian $i\cl_{\rm ext}(t)$ in terms of  the external force will be specified later.

It should be noted that the Liouvillian is not self-adjoint for dissipative systems. 
The adjoint Liouvillian is defined by the equation of the phase function or the $N-$body distribution function 
$\rho(\Vect{\Gamma},t)$:
\begin{equation}\label{rho}
\rho(\Vect{\Gamma},t)=
\tilde{U}_\leftarrow(t,0)\rho(\Vect{\Gamma},0), {~~}
\frac{\partial \rho(\Vect{\Gamma},t)}{\partial t}=-i{\cl}^\dagger(t) \rho(\Vect{\Gamma},t),
\end{equation}
where $
\tilde{U}_\leftarrow(t,0)=T_\leftarrow e^{-i \int_{0}^tds \cl^\dagger(s)}\equiv \sum_{n=0}^{\infty}
(-)^n\int_{0}^t ds_1 \cdots \int_{0}^{s_{n-1}}ds_n i\cl^\dagger(s_1) \cdots i \cl^\dagger(s_n)$
is defined by the left time ordered integral.
% \nonumber\\
 %&\equiv&
%\sum_{n=0}^{0}(-)^n\int_{0}^t ds_1\int_{0}^{s_1}ds_2\cdots \int_{0}^{s_{n-1}}ds_n i\cl^\dagger(s_1) i\cl^\dagger(s_2)\cdots i \cl^\dagger(s_n) .
%\end{eqnarray}
The adjoint Liouvillian satisfies
\begin{equation}\label{eq:17}
i\cl^\dagger(t)=i{\cl}(t)+\Lambda(\Vect{\Gamma}) ,
\end{equation}
where 
%\begin{equation}\label{Lambda}
$\Lambda(\Vect{\Gamma})\equiv \frac{\partial}{\partial \bs{\Gamma}} \cdot \dot{\bs{\Gamma}}(\bs{\Gamma}) =\sum_{i} \frac{\partial}{\partial \bs{p}_{i}} \cdot \bs{F}_{i}^{\rm (vis)}$
%\end{equation} 
is the phase volume contraction.  
Note that $\Lambda(\bs{\Gamma})$ is not an operator but merely a number in classical situations.
We also note that $\Lambda(\bs{\Gamma})$ is directly related to the Jacobian as $|\partial \bs{\Gamma}(t)/\partial \bs{\Gamma}|=\exp[\int_0^td\tau\Lambda(\bs{\Gamma}(\tau))]$.
Equation (\ref{eq:17}) implies that non-Hermitian part appears only through $\Lambda(\bs{\Gamma})$. This means that $i\cl_{\rm ext}(\bs{\Gamma},t)$ should be a Hermitian as in the case of $i\cl ^{\rm (el)}(\bs{\Gamma})$.

The average of a physical quantity $A(\bs{\Gamma}(t))$ is defined as
\begin{equation}\label{average}
\langle A(\bs{\Gamma}(t))\rangle \equiv \int d\Vect{\Gamma} \rho(\Vect{\Gamma},0)A(\Vect{\Gamma}(t))=\int d\Vect{\Gamma} A(\Vect{\Gamma})\rho(\Vect{\Gamma},t) .
\end{equation}
From Eqs. (\ref{Liouville}), (\ref{rho}) and (\ref{average}) we obtain the relations
\begin{equation}\label{average2}
\int d\Vect{\Gamma} \rho(\Vect{\Gamma})U_\rightarrow(0,t)A(\Vect{\Gamma})=\int d\Vect{\Gamma} A(\Vect{\Gamma})
\tilde{U}_\leftarrow(t,0)\rho(\Vect{\Gamma},0)
\end{equation}
and
\begin{equation}\label{average3}
\int d\Vect{\Gamma} \rho(\Vect{\Gamma})i\cl(t) A(\Vect{\Gamma})=-\int d\Vect{\Gamma} A(\Vect{\Gamma}) i\cl^\dagger(t) \rho(\Vect{\Gamma},0) .
\end{equation}
%Note that Eq. (\ref{average}) can be used for any $t$.
% and does not have to assume the starting point from $t=0$ as in Eq.(\ref{average2})by replacing $\bs{\Gamma}$ by $\bs{\Gamma}(t_0)$. 
We also have the following relation for the inverse path:
\begin{eqnarray}\label{claasical-13-b}
\langle \check{A}(\bs{\Gamma}(-t)) \rangle
&=&\int d\mathbf{\Gamma} \rho(\mathbf{\Gamma},0) \{ U_{\leftarrow}(t,0)A(\mathbf{\Gamma}) \}
\nonumber\\
&=&\int d\mathbf{\Gamma}  A(\mathbf{\Gamma}) \{ \tilde{U}_\rightarrow (0,t)\rho(\mathbf{\Gamma},0)\} ,
\end{eqnarray}
where $\check{A}(\bs{\Gamma}(-t))%=U_\leftarrow(t,0)A(\bs{\Gamma})
=U_\leftarrow(t,0)A(\bs{\Gamma})U_\rightarrow(0,t)$
and $\check{A}(\bs{\Gamma}(t))=A(\bs{\Gamma}(t))=U_\rightarrow(0,t)A(\bs{\Gamma})U_\leftarrow(t,0)$ for $t\ge 0$.
Here, we have introduced $U_\leftarrow(t,0)\equiv T_\leftarrow \exp[-i\int_0^tds \cl (s)]$.
Note that $\check{A}(\bs{\Gamma}(-t))$ is not equal to $A(\bs{\Gamma}(-t))$ except for the case that the system has time translational symmetry.
Therefore,  
$U_\leftarrow(0,-t)=U_\rightarrow^{-1}(-t,0)$ and $\tilde{U}_\rightarrow(-t,0)=\tilde{U}_\leftarrow^{-1}(0,-t)$ are, respectively, not equal to
$U_\leftarrow(t,0)=U_\rightarrow^{-1}(0,t)$ and $\tilde{U}_\rightarrow(0,t)=\tilde{U}_\leftarrow^{-1}(t,0)$ in general,
where $\tilde{U}_\rightarrow(0,t)\equiv T_\rightarrow \exp[i\int_0^t ds \cl^\dagger(s) ]$.
In other words, the useful relations $U_\leftarrow(t,0)=U_\leftarrow(0,-t)$ and $\tilde{U}_\rightarrow(-t,0)=\tilde{U}_\rightarrow(0,t)$ can be used, if the Liouville operator is independent of time.

In the last part of this section, we introduce a useful formula between $\tilde{U}_\rightarrow(0,t)$ and $U_\rightarrow(0,t)$.
It is easy to show that  two evolution operators $U_\rightarrow(0,t)$ and $\tilde{U}_\rightarrow(0,t)$  satisfy Dyson's equation \cite{Evans08}:
\begin{equation}\label{dyson}
\tilde{U}_\rightarrow(0,\tau)=U_\rightarrow(0,\tau)+\int_{0}^{\tau}ds \tilde{U}_\rightarrow(0,s)
\Lambda(\Vect{\Gamma})
U_\rightarrow(s,\tau) ,
\end{equation}
where we have used Eq. (\ref{eq:17}). It is straightforward to rewrite Eq. (\ref{dyson}) as \cite{Evans08,kawasaki}
\begin{equation}\label{kawasaki}
\tilde{U}_\rightarrow(0,t)=\exp
\left[\int_0^td\tau \Lambda(\bs{\Gamma}(\tau))  \right]
U_\rightarrow(0,t) ,
\end{equation}
where $\Lambda(\bs{\Gamma}(t))=U_\rightarrow(0,t)\Lambda(\bs{\Gamma})U_\leftarrow(t,0)$.
From the inversion relation $\tilde{U}_\leftarrow(t,0)=\tilde{U}_\rightarrow^{-1}(0,t)$, we also have another identity
\begin{equation}\label{inverse_U}
\tilde{U}_\leftarrow(t,0)=U_\leftarrow(t,0)\exp\left[-\int_0^td\tau \Lambda(\bs{\Gamma}(\tau))\right] .
\end{equation}

\section{Some nonequilibrium identities}

In this section, we derive some nonequilibrium identities for dissipative classical systems.
First, we introduce IFT in Sec. III A which is the consequence of the conservation of the probability as well as another derivation of IFT in Appendix A.
We also introduce a quantity which corresponds to the entropy in our system.
Second, we derive the standard fluctuation theorem for the dissipation rate from IFT (Sec. III B).
Third, we extend the IFT to the Jarzynski equality if the dynamics includes a part not described by the Liouvillian (Sec. III C) as well as its derivation starting from the canonical state in Appendix B.
Fourth, we  discuss the generalized Green-Kubo formula as an expression of nonlinear response theory for the steady dynamics (Sec. III D).
Note that  the effect of measurement can easily be included in our formulation, which will be discussed in Appendix C.

\subsection{Integral Fluctuation Theorem}

Let us consider the time evolution from a state characterized by
\begin{equation}
\rho(\Vect{\Gamma},0)=\rho_{\rm NE}(\Vect{\Gamma})\equiv \frac{e^{- I(\mathbf{\Gamma})}}{\cal Z} \quad,
\label{large-deviation}
\end{equation} 
where $I(\Vect{\Gamma})$ is the effective potential in a nonequilibrium state, and 
${\cal Z}\equiv \int d\Vect{\Gamma}e^{-I(\mathbf{\Gamma})}
%=\int d\Vect{\Gamma}(t) e^{-I(\bs{\Gamma}(t))}
$.
There are two advantages to introduce $\rho_{\rm NE}(\Vect{\Gamma})$ instead of using the equilibrium distribution function:
(i) We can demonstrate that the fluctuation theorem and related identities are independent of the choice of a reference state, and
(ii) we can discuss the response theory around the nonequilibrium state characterized by $\rho_{\rm NE}(\bs{\Gamma})$.

The normalization factor ${\cal Z}$ is invariant if the time evolution is only described by the Liouvillian. 
Namely, we can rewrite ${\cal Z}=\int d\bs{\Gamma}(t) e^{-I(\bs{\Gamma}(t))}$. 
From these two expressions for ${\cal Z}$ with the aid of $d\bs{\Gamma}(t)=d\bs{\Gamma} e^{\int_0^t\Lambda(\bs{\Gamma}(\tau))} $ we obtain
\begin{equation}\label{IFT}
\langle e^{-\int _0^t d\tau \Omega(\bs{\Gamma}(\tau))} \rangle_{\rm NE}=1 ,
\end{equation}
where 
\begin{equation}\label{Omega}
\Omega(\bs{\Gamma}(t))=\dot{I}(\bs{\Gamma}(t))-\Lambda(\bs{\Gamma}(t))
\end{equation}
and $\langle A(\bs{\Gamma}) \rangle_{\rm NE}\equiv \int d\bs{\Gamma} \rho_{\rm NE}(\bs{\Gamma}) A(\bs{\Gamma}) $ for an arbitrary function $A(\bs{\Gamma})$, 
and $\dot{I}(\bs{\Gamma}(t))=(d/dt)I(\bs{\Gamma}(t))$.
Here, we have used that $I(\bs{\Gamma}(t))$ is commutable with $\Lambda(\bs{\Gamma}(t))$. 
Equation (\ref{IFT}) is IFT~\cite{Seifert12,Seifert05}, but the essence of IFT is the conservation law of the probability which is independent of the details for the dynamics.
Note that $\Omega(\bs{\Gamma}(t))$ introduced here is an extension of the dissipation function or a generalized entropy production in thermostat systems~\cite{Evans08}.
We present another derivation of IFT in Appendix A.

With the aid of Janssen's inequality and its differentiation with respect to the time, we readily obtain an inequality similar to the second law of thermodynamics:
\begin{equation}\label{inequality}
\langle \Omega(\bs{\Gamma}(t) ) \rangle_{\rm NE} \ge 0 ,
\end{equation}
which determines the stability of the state.
This is a quite natural result, because $\Omega(\bs{\Gamma}(t))$ corresponds to a generalized entropy production rate~\cite{Evans08}.
According to (\ref{inequality}) we can introduce the entropy-like quantity
\begin{equation}\label{entropy}
S(\bs{\Gamma}(t))\equiv \int_0^td\tau \langle \Omega(\bs{\Gamma}(\tau)) \rangle_{\rm NE},
\end{equation}
which increases with time.
Further physical implication of the entropy-like quantity (20) will be discussed in Sec. V.

Equation (\ref{IFT}) is the exact relation, but is not easy to be verified from experiments and simulations because of the limitation of accuracy for $N-$body correlation function.
Indeed, Ref. \cite{Evans90} stated that the normalization from the conservation of probability is difficult to be achieved by numerical simulations.
Instead, we can use an alternative expression
\begin{equation}\label{omega(t)}
\omega(t)\equiv \left\langle \Omega(\bs{\Gamma}(t))\exp\left[\int_0^td\tau \Omega(\bs{\Gamma}(\tau))\right]\right\rangle_{\rm NE}=0 
\end{equation}
for the practical purpose, which is obtained from the time derivative of Eq. (\ref{IFT}).
The relation (\ref{omega(t)}) is highly nontrivial. Indeed if we adopt the decoupling approximation for $\omega(t)$, it should satisfy
$\omega(t)\simeq \langle \Omega(\bs{\Gamma}(t)) \rangle_{\rm NE}/\langle e^{-\int_0^t\Omega(\bs{\Gamma}(\tau))} \rangle_{\rm NE} \ge 0$
thanks to Eqs. (\ref{IFT}) and (\ref{inequality}).
Therefore, the numerical verification of Eq. (\ref{omega(t)}) which is the result of the correlation effects is important for $\Omega(\bs{\Gamma}(t))\ne 0$.

%\end{eqnarray}

\subsection{Fluctuation theorem}

It is straightforward to derive the conventional fluctuation theorem (FT) from IFT (\ref{IFT}), where FT is the relation of the probability of the entropy production  between the forward path and the inverse path \cite{Evans08}.
Now, let us derive the conventional fluctuation theorem from the integral fluctuation theorem (\ref{IFT}).
% under the steady dynamics 
%In this case, we can use the time translational symmetry within the evolution $i\cl_+$ or $i\cl_-$. 
We, here, assume that the initial distribution is given by 
$\rho_{\rm NE}(\bs{\Gamma})=\rho_{\rm eq}(\bs{\Gamma})$:
\begin{equation}\label{eq:rho_eq}
\rho_{\rm eq}(\bs{\Gamma})=\frac{e^{-\beta H(\bs{\Gamma})}}{Z(\beta)};\quad H(\bs{\Gamma})=\sum_i\frac{\bs{p}_i^2}{2m}+
\frac{1}{2}\sum_{i,j\ne i}V(r_{ij}) 
\end{equation}
with $Z(\beta)\equiv \int d\bs{\Gamma}e^{-\beta H(\bs{\Gamma})}$ and the inverse temperature $\beta$.

Now let us consider the process from time 0 to time $t$ by the time evolution operator $U_\rightarrow(0,t)$ and the trajectory of the phase variable $\bs{\Gamma}(\tau)=U_\rightarrow(0,\tau)\bs{\Gamma}$ for $0\le \tau \le t$.
The inverse process, thus, is characterized by the time evolution operator $U_\leftarrow(t,0)$ and the 
 inverse phase variable 
$\bs{\Gamma}^*(\tau)\equiv \{ \bs{r}_i(t-\tau), -\bs{p}_i(t-\tau) \}_{i=1}^N=\{\bs{\Gamma}(t-\tau)\}^T$ for $0\le \tau \le t$,
 where the operation $\{\bs{\Gamma}(t) \}^T$ represents the change of the sign of the momenta
 $\{\bs{\Gamma}(t) \}^T\equiv \{\bs{r}_i(t), -\bs{p}_i(t) \}_{i=1}^N$. 
Because the probability of the inverse trajectories $\rho_R(\bs{\Gamma}^*(0))
=\rho_{\rm eq}(\bs{\Gamma}^*)
$
 is still normalized as $\int d\bs{\Gamma}^* \rho_{\rm eq}(\bs{\Gamma}^*)=1$ with the abbreviation $\bs{\Gamma}^*\equiv \bs{\Gamma}^*(0)$, 
Eq. (\ref{IFT}) can be rewritten as
\begin{equation}
\int d\bs{\Gamma} \rho_{\rm eq}(\bs{\Gamma}) e^{-t \overline{\Omega_t}}
=\int d\bs{\Gamma}^* \rho_{\rm eq}(\bs{\Gamma}^*) ,
\label{FT_steady2}
\end{equation}
where we have introduced $\overline{\Omega_t}\equiv \frac{1}{t}\int_0^t d\tau \Omega_{\rm eq}(\bs{\Gamma}(\tau))$
with $ \Omega_{\rm eq}(\bs{\Gamma}(t))=\beta \dot{H}(\bs{\Gamma}(t))-\Lambda(\bs{\Gamma}(t)) $.
%With the aid of the time reversal symmetry for $\rho_{\rm eq}(\bs{\Gamma})=\rho_{\rm eq}(\bs{\Gamma}^T)$ and $d\bs{\Gamma}^T=d\bs{\Gamma}|\partial \bs{\Gamma}^T/\partial \bs{\Gamma}|$, 
%Eq. (\ref{FT_steady2}) means that the Jacobian $|\partial \bs{\Gamma}^T/\partial \bs{\Gamma}|$ is given by $\exp[-t\overline{\Omega}_t]$.
%Thus, the integrand of Eq.(\ref{FT_steady2}) satisfies $d\bs{\Gamma} \rho_{\rm eq}(\bs{\Gamma}) e^{-t \overline{\Omega_t}}
%=d\bs{\Gamma}^T \rho_{\rm eq}(\bs{\Gamma}^T)$. 
From the definition of $\Lambda(\bs{\Gamma}(t))$ and the Hamiltoninan, there are some trivial relations: 
$\Lambda(\bs{\Gamma}^*(\tau))=-\Lambda(\bs{\Gamma}(t-\tau))$,
$H(\bs{\Gamma}^*(\tau))=H(\bs{\Gamma}(t-\tau))$ and
$\Omega_{\rm eq}(\bs{\Gamma}^*(\tau))=-\Omega_{\rm eq}(\bs{\Gamma}(t-\tau))$ for $0\le \tau \le t$. 
%Here, $\bs{\Gamma}_+(\pm t)=e^{\pm i\cl_+t}\bs{\Gamma}$.
Therefore, we can write the probability of $\overline{\tilde{\Omega}_t}=-A$ for 
%$\tilde{\Omega}_{\rm eq}(\bs{\Gamma})\equiv \Omega_{\rm eq}(\bs{\Gamma}^T)$ and 
$\overline{\tilde{\Omega}_t}\equiv \frac{1}{t}\int_0^td\tau {\Omega}_{\rm eq}(\bs{\Gamma}^*(\tau))
=\frac{1}{t}\int_0^t d\tau \Omega_{\rm eq}(\{\bs{\Gamma}(\tau)\}^T) $: 
\begin{eqnarray}\label{conventional_FT}
{\rm Prob}(\overline{\tilde{\Omega}_t}=-A)
&=&\int d\bs{\Gamma}^* \rho_{\rm eq}(\bs{\Gamma}^*)\delta(\overline{\tilde{\Omega}_t}+A)
\nonumber\\
&=& \int d\bs{\Gamma}(t) \rho_{\rm eq}(\bs{\Gamma}(t))\delta(\overline{{\Omega}_t}-A) \nonumber\\
&=& \int d\bs{\Gamma} \rho_{\rm eq}(\bs{\Gamma})e^{-t\overline{\Omega}_t}\delta(\overline{\Omega_t}-A)
\nonumber\\
&=& e^{-At}\int d\bs{\Gamma}\rho_{\rm eq}(\bs{\Gamma})\delta(\overline{\Omega_t}-A)
\nonumber\\
&=& e^{-At} {\rm Prob}(\overline{\Omega_t}=A) 
\end{eqnarray}
for the conventional fluctuation theorem, where we have used $|\partial \bs{\Gamma}^*(\tau)/\partial \bs{\Gamma}(t-\tau)|=1$,
$\rho_{\rm eq}(\bs{\Gamma}(t))=\rho_{\rm eq}(\bs{\Gamma})e^{-\beta \int_0^t d\tau \dot{H}(\tau)}$
and $ d\bs{\Gamma}(t)= d\bs{\Gamma}e^{\int_0^t d\tau \Lambda(\bs{\Gamma}(\tau))}$.
Note that  the argument is still valid for the general starting point Eq. (\ref{large-deviation}) if we have the symmetry $I(\bs{\Gamma}^*(\tau))=I(\bs{\Gamma}(t-\tau))$. 

%Note that our fluctuation theorem 
%\begin{equation}\label{form_FT}
%{\rm Prob}(\overline{\tilde{\Omega}_t}=A)=e^{-At} {\rm Prob}(\overline{\Omega_t}=A) 
%\end{equation}
%differs from the standard form ${\rm Prob}(\overline{\tilde{\Omega}_t}=-A)=e^{-At}{\rm Prob}(\overline{\Omega_t}=A)$, because our system does not have the time reversal symmetry.

It is easy to verify that IFT \eqref{IFT} 
can be derived from the fluctuation theorem (FT) 
\eqref{conventional_FT}. 
However, it should be noted that FT is the expression 
between the real trajectory and its inverse trajectory. 
Although there exists the inverse trajectory even 
for any dissipative systems, 
this trajectory cannot be traced by any physical operation. 
In this sense, FT \eqref{conventional_FT} is useless, 
but IFT \eqref{conventional_FT} still has the physical relevancy.
Nevertheless, as in recent experiments~\cite{joubaud12,Naert12,mounier12} if we use an asymmetric rotor with some vanes in a granular gas, it might be possible to observe FT presented here.

\subsection{Jarzynski equality}

%THIS SECTION IS STILL UNCLEAR. IN WHAT SITUATION WE NEED THE TIME DEPENDENCE OF THE NORMALIZATION?

Jarzynski equality is an identity which is related to the second law of thermodynamics \cite{Jarzynski}.
The integral fluctuation theorem (\ref{IFT}) can be regarded as one of the expressions of Jarzynski equality, but Jarzynski equality in a narrow sense involves the change of the normalization during the process we consider.
It should be noted that if the time evolution is only governed by the Liouville operator, there is no room that ${\cal Z}$ is changed.
In other words, we should consider the dynamics  which cannot be characterized by the Liouvillian such as the change of the system volume.

In this subsection, we demonstrate the existence of Jarzynski equality starting from Eq.(\ref{large-deviation}) with an abstract way.
More instructive proof of Jarzynski equality starting from Eq. (\ref{eq:rho_eq}) is presented in Appendix B, in which the change of energy in the system is connected with the work acting on the system.

To describe such a process let us introduce a protocol parameter $\lambda(\tau)$ for $0\le \tau \le t$ which has the value $\lambda(0)=0$ at the initial instance and $\lambda(t)=1$ at time $t$.
As  the Hamiltonian changes by the external work in the original argument by Jarzynski~\cite{Jarzynski}, 
the effective potential changes during the operation from $I(\bs{\Gamma},0)$ to $I(\bs{\Gamma}(t),1)$
~\footnote{We may be able to prepare the initial $I(\bs{\Gamma},0)$ without the introduction of the protocol parameter, i.e. $\lambda(\tau)=0$ for $\tau<0$.
}.
As a result, the normalization ${\cal Z}(0)=\int d\bs{\Gamma} e^{-I(\bs{\Gamma},0)}$ at the initial instance can become 
${\cal Z}(t)=\int d\bs{\Gamma}(t) e^{-I(\bs{\Gamma}(t),1)}$.
If we set ${\cal Z}(t)=e^{-{\cal F}(t)}$, we can write
\begin{eqnarray}\label{jarzynski}
e^{-\Delta {\cal F}}
&=& \frac{{\cal Z}(t)}{{\cal Z}(0)}
=\frac{1}{{\cal Z}(0)}\int d\bs{\Gamma}(t) e^{-I(\bs{\Gamma}(t),1)}
\nonumber\\
&=& \int d\bs{\Gamma} \left|\frac{\partial{\bs{\Gamma}(t)}}{\partial \bs{\Gamma}} \right|\frac{e^{-I(\bs{\Gamma},0)}}{{\cal Z}(0)}
e^{-\int_0^t d\tau \dot{I}(\bs{\Gamma}(\tau),\lambda(\tau))}
\nonumber\\
&=& \left\langle \exp\left[-\int_0^t d\tau \Omega(\bs{\Gamma}(\tau),\lambda(\tau)) \right]  \right\rangle_{\rm NE} ,
\end{eqnarray}
where $\Delta {\cal F}\equiv {\cal F}(t)-{\cal F}(0)$.
Equation (\ref{jarzynski}) is the Jarzynski equality for dissipative systems.

We should note that ${\cal Z}(t)$ still has the meaning of the partition function under a nonequilibrium weight function $I(\bs{\Gamma}(t),1)$.
Therefore, we can regard ${\cal F}(t)$ as the effective free energy for dissipative system at $t$.
If we start from $\rho_{\rm eq}(\bs{\Gamma})$ it is straightforward to show the Jarzynski equality is reduced to
$e^{-\beta \Delta F}= \langle \exp [ -\int_0^td\tau \Omega(\bs{\Gamma}(\tau),\lambda(\tau))] \rangle_{\rm eq}$, 
where $\Delta F=F(t)-F(0)$ with $e^{-\beta F(t)}=  Z(t)=\int d\bs{\Gamma}(t) e^{-\beta H(t)}$.
Even in this case, we have to take into account the phase volume contraction $\Lambda(\bs{\Gamma}(t))$ for dissipative systems.

%Here, we briefly indicate that IFT introduced in the previous subsection is directly connected with Jarzynski equality if we assume that the Hamiltonian depends on time explicitly.

\subsection{generalized Green-Kubo formula}

%For $t>0$ Eq. (\ref{average}) can be written as
%\begin{eqnarray}\label{<A(t)>}
%\langle A(\bs{\Gamma}(t)) \rangle_{\rm NE}
%&=&
%\int d\bs{\Gamma}A(\bs{\Gamma}) U_\leftarrow(t,0)
%\nonumber\\
%&& \times
%\left[e^{-\int_0^td\tau \Lambda(\bs{\Gamma}(\tau))}\rho_{\rm NE}(\bs{\Gamma})
%\right] ,
%\end{eqnarray}
%where we have used Eq. (\ref{inverse_U}).
%Unfortunately, we cannot get any useful relations from Eq. (\ref{<A(t)>}).

%In contrast to the previous case, unfortunately, we cannot obtain a useful relation from Eq. (\ref{<A(t)>}).
%Unfortunately, it is difficult to obtain a meaningful relation from Eq. (\ref{<A(t)>}). 
The generalized Green-Kubo formula is used for the response around an equilibrium state under a nonequilibrium steady external force.~\cite{Evans08}
Unfortunately, this relation cannot be proven under non-stationary external forces, 
but is valid, as will be shown, for the response around a nonequilibrium steady state under a sudden change of the external force. 

Now, let us restrict our interest to the steady dynamics:
\begin{equation}\label{caseA}
i\cl(t)=\begin{cases}
i \cl_+ & \text{for $t\ge 0$} \\
i \cl_- & \text{for $t<0$} ,
\end{cases}
\end{equation}
where $i\cl_+$ and $i\cl_-$ are respectively steady Liouville operators for the generalized Green-Kubo formula.
 %\eqref{caseA}.
Therefore, we can replace Eq. (\ref{claasical-13-b}) by
\begin{eqnarray}
\langle A(\bs{\Gamma}_+(t)) \rangle_{\rm NE}&=&\int d\bs{\Gamma}\rho_{\rm NE}(\bs{\Gamma})e^{i\cl_+ t}A(\bs{\Gamma})
\nonumber\\
&=&\int d\bs{\Gamma}A(\bs{\Gamma})e^{-i\cl_+^\dagger t}\rho_{\rm NE}(\bs{\Gamma}) .
\label{A(Gamma_-(t)}
\end{eqnarray}
Thus, we can obtain a compact expression for the distribution function: \cite{Evans08,Chong09b}
\begin{equation}\label{distribution_at_t}
\rho(\bs{\Gamma},t)=\exp\left[\int_0^td\tau\Omega(\bs{\Gamma}_+(-\tau))\right]\rho_{\rm NE}(\bs{\Gamma}) ,
\end{equation}
where $\bs{\Gamma}_+(\pm\tau)=e^{\pm i\cl_+ \tau}\bs{\Gamma}$.
%\footnote{
The derivation of Eq. (\ref{distribution_at_t}) is simple. 
The inverse of Eq. (\ref{rho}) with the help of Eqs. (\ref{kawasaki}) and (\ref{large-deviation}) is given by
$\rho_{\rm NE}(\bs{\Gamma})=e^{i\cl^\dagger_+ t}\rho(\bs{\Gamma}_+,t)=e^{\int_0^td\tau\Lambda(\bs{\Gamma}_+(\tau))}\rho(\bs{\Gamma}_+(t),t)$.
Let us operate $e^{-i\cl_+t}$ on the both sides. 
The right hand side on this equation can be rewritten as
$e^{-i\cl_+ t}[
e^{\int_0^td\tau\Lambda(\bs{\Gamma}_+(t-\tau))}\rho(\bs{\Gamma}_+(t),t)]
=e^{\int_0^td\tau \Lambda(\bs{\Gamma}_+(-\tau))}\rho(\bs{\Gamma}_+,t)$, while the left hand side is replaced by
$\rho(\bs{\Gamma}_+(-t))=e^{-{I}(\bs{\Gamma}_+(-t))}/{\cal Z}=e^{\int_0^td\tau \dot{{I}}(\bs{\Gamma}_+(-\tau))}\rho_{\rm NE}(\bs{\Gamma})$, where $\dot{{I}}(\bs{\Gamma}_+(-\tau))=d{I}(\bs{\Gamma}_+(-\tau))/d\tau'$ with $\tau'=-\tau$.
It is obvious that the derivation assumes the translational invariance of the Liouville operator
\footnote{We should note that $\rho(\bs{\Gamma},-t)=\exp[-\int_0^td\tau \Omega(\bs{\Gamma}(\tau))]\rho_{\rm NE}(\bs{\Gamma})$ is held from the second derivation of (\ref{IFT}).
\label{rho(-)}
}.

%}

With the aid of Eq.(\ref{distribution_at_t}), the time derivative of Eq. (\ref{A(Gamma_-(t)}) is reduced to
\begin{eqnarray}\label{Green-Kubo}
\frac{d}{dt}\langle A(\bs{\Gamma}_+(t)) \rangle_{\rm NE}
&=& \int d\bs{\Gamma}_+ A(\bs{\Gamma})\Omega(\bs{\Gamma}_+(-t))\rho(\bs{\Gamma}_+,t)
\nonumber\\
&=& \int d\bs{\Gamma}_+ e^{i\cl_+t}[A(\bs{\Gamma})\Omega(\bs{\Gamma}_+(-t))]\rho_{\rm NE}(\bs{\Gamma})
\nonumber\\
&=&
\langle A(\bs{\Gamma}_+(t))\Omega(\bs{\Gamma}) \rangle_{\rm NE}.
\end{eqnarray}
Its integral form, then, becomes
\begin{equation}\label{Green-Kubo2}
\langle A(\bs{\Gamma}_+(t)) \rangle_{\rm NE}
=\langle A(\bs{\Gamma}) \rangle_{\rm NE} +\int_0^t d\tau \langle A(\bs{\Gamma}_+(t))\Omega(\bs{\Gamma}) \rangle_{\rm NE} .
\end{equation}
Equation (\ref{Green-Kubo2}) states that the generalized Green-Kubo formula in Refs. \cite{Evans08,Chong09b} is still valid even when we start from an arbitrary nonequilibrium state under the steady dynamics given by Eq. (\ref{caseA}).
Note that Eq. (\ref{Green-Kubo2}) is reduced to the well-known Green-Kubo formula if we start from the equilibrium state in the zero dissipation limit~\cite{Chong09b}.

\section{Application to sheared granular fluids}

So far, our formulation is so general that we can apply any classical dissipative systems described by the Liouville equation, though the effect of boundaries is not clear.
In this section, we verify the validity of the identities we obtain in the previous section from the direct simulation of sheared frictionless granular particles.

\subsection{Setup for sheared granular fluid}
 
 Let us consider a system of $N$ smooth granular particles, where the rotation and the tangential contact force of particles can be ignored.
The interaction between particles, thus, is assumed to be characterized only by the normal contact force for simplicity.  
We also assume that a sheared system
can be described by the SLLOD equations~\cite{Evans08}
\begin{subequations}
\label{eq:SLLOD}
\begin{eqnarray}
\dot{\Vect{r}}_{i} &=& \frac{\Vect{p}_{i}}{m} + {\sf \kappa}(t) \cdot \Vect{r}_{i},
\label{eq:SLLOD-a}
\\
\dot{\Vect{p}}_{i} &=& \Vect{F}_{i}^{\rm (el)} + \Vect{F}_{i}^{\rm (vis)} - {\sf \kappa}(t) \cdot \Vect{p}_{i} ,
\label{eq:SLLOD-b}
\end{eqnarray}
\end{subequations}
where 
$\dot{\Vect{r}}_i$ refers to the velocity of $i$th particle,  
$\Vect{p}_{i}$ is the peculiar momentum 
defined by Eq. (\ref{eq:SLLOD-a})  and ${\sf \kappa}(t)$ is  the shear rate tensor whose component is denoted by $\kappa_{\alpha\beta}(t)$.
Here, the force acting on $i$th particle consists of two parts, i.e. 
$\Vect{F}_i=\Vect{F}_i^{\rm (el)}+\Vect{F}_i^{\rm (vis)}$.
The conservative force %$\Vect{F}_{i}^{\rm (el)} = - \partial U / \partial \Vect{r}_{i}$ with the total interaction potential $U$ 
is given by a sum $\Vect{F}_{i}^{\rm (el)} = \sum_{j \ne i} \Vect{F}_{ij}^{\rm (el)}$
of the elastic repulsive forces exerted on the $i$th particle by the other $j$th particle: 
\begin{equation}
\Vect{F}_{ij}^{\rm (el)} = %- \frac{\partial V(r_{ij})}{\partial \Vect{r}_{ij}} =
\Theta(d-r_{ij}) f(d-r_{ij}) \hat{\Vect{r}}_{ij},
\end{equation}
where $d$ is the diameter of the particles, and
%$V(r)$ is the pair potential; 
%$\Vect{r}_{ij} = \Vect{r}_{i} - \Vect{r}_{j}$, $r_{ij} = | \Vect{r}_{ij} |$, $\hat{\Vect{r}}_{ij} = \Vect{r}_{ij} / r_{ij}$;
%$d$ denotes the particle diameter; 
$\Theta(x)$ is the Heaviside step function, i.e. $\Theta(x)=1$ for $x>0$ and $\Theta(x)=0$ for otherwise.
In our simulation, we adopt the forms of the elastic repulsive force :
\begin{equation}
f(x) = k x
\end{equation}
with the elastic constant $k$.
Similarly, the viscous dissipative force $\Vect{F}_{i}^{\rm (vis)}$
due to inelastic collisions between particles is represented by 
a sum $\Vect{F}_{i}^{\rm (vis)} = \sum_{j \ne i} \Vect{F}_{ij}^{\rm (vis)}$ of
two-body contact forces 
\begin{equation}
\Vect{F}_{ij}^{\rm (vis)} = 
-\hat{\bs{r}}_{ij}{\cal F}(r_{ij}) (\Vect{g}_{ij} \cdot \hat{\Vect{r}}_{ij}),
%- (\Vect{g}_{ij} \cdot \hat{\Vect{r}}_{ij}) 
\label{eq:F-vis-def}
\end{equation}
where $\hat{\bs{r}}_{ij}\equiv \bs{r}_{ij}/r_{ij}$, $\bs{g}_{ij} \equiv \dot{\bs{r}}_{ij} 
=(\bs{p}_{i} - \bs{p}_{j})/m + {\sf \kappa}(t) \cdot \Vect{r}_{ij}$ and
\begin{equation}
{\cal F}(r_{ij})\equiv \Theta(d-r_{ij}) \zeta.
\end{equation}
The amount of energy dissipation upon inelastic collisions 
is characterized by the viscous constant $\zeta$.
It should be noted that Eq. (\ref{eq:SLLOD}) is reduced to
Newtonian equations
\begin{equation}\label{newton_eq}
m \ddot{\Vect{r}}_i=\Vect{F}_i ,
\end{equation}
if ${\sf \kappa}(t)$ is independent of $t$, where $\ddot{\bs{r}}_i$ denotes the acceleration of $\bs{r}_i$.
In this paper, we restrict our interest to the case of $\kappa_{\alpha\beta}(t)=\dot\gamma(t) \delta_{\alpha,x}\delta_{\beta,y}$, for simplicity, and
adopt the Lees-Edwards boundary condition to remove complicated effects  of the boundaries.

%where $\bs{F}_i^{\rm (vis)}$ is the viscous force acting on $i-$th particle which is represented by
%$\bs{F}_i^{\rm (vis)}=\sum_{j\ne i}\bs{F}_{ij}^{\rm (vis)}$.
The Liouville operator representing the shear flow $i\cl_{\rm ext}(t) =i\cl_{\dot{\gamma}(t)}$ is given by
\begin{equation}\label{L_s}
i\cl_{\dot{\gamma}(t)}(\bs{\Gamma})=\dgam\sum_{j=1}^N\left(y_j\frac{\partial}{\partial x_j}-p_{y,j}\frac{\partial}{\partial p_{x,j}}\right) .
\end{equation}

For our setup (\ref{eq:SLLOD})-(\ref{L_s}), the phase-space compression factor, $\Lambda(\bs{\Gamma}(t))$ 
which only depends on $t$ through $\bs{\Gamma}(t)$ is written as
\begin{equation}
\Lambda(\bs{\Gamma}) = \sum_{i} \frac{\partial}{\partial \bs{p}_{i}} \cdot \bs{F}_{i}^{\rm (vis)} =
- \frac{1}{m} \sum_{i} \sum_{j \ne i} {\cal F}(r_{ij})
\label{eq:SLLOD-Lambda}
\end{equation}
for $t\ge 0$.

We consider the case that
%the essentially Eq.(\ref{caseA}) starting from the canonical distribution function $\rho_{\rm eq}(\bs{\Gamma})$ in Eq.(\ref{eq:rho_eq}).
the dynamics is given by
\begin{equation}\label{caseC}
i\cl(t)=\begin{cases}
i\cl_1
 & \text{for $t\ge 0$} \\
 i\cl_0 & \text{for $-t_0 \le t<0$}  , \\
\end{cases}
\end{equation}
where $i{\cal L}_0$ and $i{\cal L}_1$ are, respectively given by $i\cl_0\equiv i \cl^{\rm (el)}+i\cl^{\rm (vis)}+i\cl_{\dot\gamma_0}$ and 
$i\cl_1\equiv  i \cl^{\rm (el)}+i\cl^{\rm (vis)}+i\cl_{\dot\gamma_1}$
with the constant shear rates $\dot\gamma_1 \ne \dot\gamma_0$,
where the initial distribution for $t < -t_0$ is assumed to be given by Eq. (\ref{eq:rho_eq}) with the inverse temperature $\beta$.
The setup at $t_0 = 0$ is used to verify the response theory from the equilibrium state (\ref{eq:rho_eq}),
while the general setup for $t_0 > 0$ is used to analyze the response from a nonequilibrium state $\rho_{\rm NE}(\bs{\Gamma})$.
It should be noted that 
 generalized Green-Kubo formula (\ref{Green-Kubo2}) is reduced to
 the known definition of the viscosity
\begin{equation}\label{GK3}
\eta=\beta V \lim_{t\to \infty}
\int_0^t d\tau \langle \sigma_{xy}(\bs{\Gamma}(\tau))\sigma_{xy}(\bs{\Gamma}) \rangle_{\rm eq}
\end{equation}
in the zero dissipation limit with $t_0 = 0$, 
where $V$ is the volume of the system \cite{Chong09b}
with the aid of $A(\bs{\Gamma}(t))=\sigma_{xy}$ which is $xy-$component of the shear stress, 
$\langle \sigma_{xy}(\bs{\Gamma}) \rangle_{\rm eq}=0$ 
and 
$\eta=\lim_{t\to \infty}\langle \sigma_{xy}(\bs{\Gamma}(t)) \rangle_{\rm eq}/\dot\gamma$.
For the calculation of the shear stress we use its microscopic expression 
\begin{equation}\label{micro_stress}
\sigma_{xy}(\bs{\Gamma})\equiv \frac{1}{V} \sum_i \left [ \frac{p_{i,x} p_{i,y}}{m}  + y_i \sum_{j\neq i}
(F_{ij,x}^{\rm (el)} + F_{ij,x}^{\rm (vis)})\right ],
\end{equation}
where $p_{i,x}$ and $F_{ij,x}$ represent the $x-$components of $\bs{p}_i$ and $\bs{F}_{ij}$, respectively.

In order to obtain the numerical configuration $\{ \bs{r}_i(t)\}_{i=1}^N$ 
of the equilibrium state at $t<-t_0$, we use the velocity rescaling method.
The momentum $\bs{p}_i(t) $ at equilibrium initial condition for $t<-t_0$ satisfies the Gaussian distribution 
$(\beta/2\pi m)^{3/2} \exp[-\beta \bs{p}_i^2/(2m)]$.

Our system satisfies 
Eq. \eqref{distribution_at_t} at $t=0$ for the distribution function 
with the replacement
$\int_0^td\tau\Omega(\bs{\Gamma}_+(-\tau))$ and 
$\rho_{\rm NE}(\bs{\Gamma})$
by
$\int_{-t_0}^0d\tau\Omega_{\rm eq}(\bs{\Gamma}_0(-t_0-\tau))=\int_0^{t_0} d\tau \Omega_{\rm eq}(\bs{\Gamma}_0(-\tau))$ and 
$\rho_{\rm eq}(\bs{\Gamma})$
with $\bs{\Gamma}_0(\pm t)=e^{\pm i\cl_0t}\bs{\Gamma}$.
Therefore, $I(\bs{\Gamma})$ in the distribution function \eqref{large-deviation}
is give by
\begin{equation}
I(\bs{\Gamma}) = \beta H(\bs{\Gamma}) -\int_0^{t_0}d\tau \Omega_{\rm eq}(\bs{\Gamma}_0(-\tau)).
\end{equation}
In the sheared system, $\Omega_{\rm eq}(\bs{\Gamma})$ is reduced to
\begin{equation}
\Omega_{\rm eq}(\bs{\Gamma}) = -\beta V \dot \gamma_0 \sigma_{xy}(\bs{\Gamma})
 -2\beta R(\bs{\Gamma})
 -\Lambda(\bs{\Gamma})
\end{equation}
with Rayleigh's dissipation function
\begin{equation}
R(\bs{\Gamma}) = \frac{\zeta}{4} \sum_i \sum_{j\neq i} \Theta(d-r_{ij})
(\bs{g}_{ij} \cdot \bs{\hat r}_{ij} )^2.
\end{equation}

It should be noted that  the normalization ${\cal Z}$ during the dynamics is unchanged because the dynamics does not include any protocol parameter.
Therefore, we can only verify IFT and the generalized Green-Kubo formula through the simulation.

\subsection{The verification of IFT}

In this subsection, we verify the validity of IFT (\ref{IFT}) or (\ref{omega(t)}) through the three-dimensional discrete element method (DEM).
As stated in the text, shear is applied to $x$ direction, and thus, $y-$dependence of the $x$ component of the velocity field is affected by the presence of shear.
Because of the numerical difficulty of the check of Eq. (\ref{IFT}), as stated in the previous section, we examine whether Eq. (\ref{omega(t)}) is valid.
In our simulation, 
we adopt the viscous constant $\zeta = 0.00045 \sqrt{mk}$, which
corresponds to the restitution coefficient $e = 0.999$.
The number of the particles is $N=18$. 
The volume fraction is $\phi = 0.66$, which is above the jamming point for a sufficiently large system.
The temperature of the equilibrium state at $t \le -t_0$ is $\beta^{-1} = 0.004 k d^2$.
We use the initial and the secondary shear rates $\dot\gamma_0=0.1 \sqrt{k/m}$ and $\dot\gamma_1=0.2 \sqrt{k/m}$, respectively.
We adopt the leapfrog algorithm with the time interval $\Delta t = 0.003 \sqrt{m/k}$.
In order to obtain $\Omega(\Gamma(t))$ numerically,
we approximate $\dot I(\Gamma(t))$ in Eq. \eqref{Omega} as
$\dot I(\Gamma(t)) \simeq \{ I(\Gamma(t+\Delta \tau)) - I(\Gamma(t))\} / \Delta \tau$ with $\Delta \tau = 0.00001\sqrt{m/k}$.

\begin{figure}
  \includegraphics[height=15em]{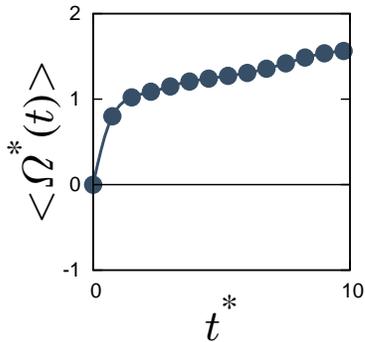}
  \caption{(Color online) The numerical verification of the inequality Eq. \eqref{inequality} for $t_0 = 0.0$. 
  Here, $\left < \Omega^*(t)\right> = \left < \Omega(t)\right>_{\rm NE} / \sqrt{k/m}$ is plotted as a function of the scaled time $t* =  t / \sqrt{m/k}$.
  }
  \label{fig1}
\end{figure}

\begin{figure}
  \includegraphics[height=15em]{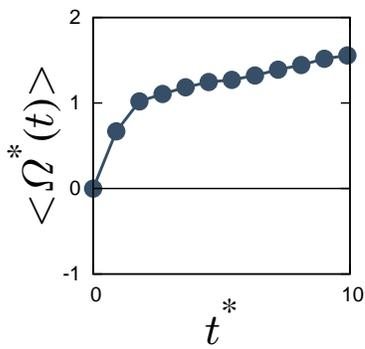}
  \caption{(Color online) The numerical verification of the inequality Eq. \eqref{inequality} for $t_0 = 1.0 \sqrt{k/m}$. 
  Here, $\left < \Omega^*(t)\right> = \left < \Omega(t)\right>_{\rm NE}/ \sqrt{k/m}$ is plotted as a function of the scaled time $t* =  t /\sqrt{m/k}$.
  }
  \label{fig2}
\end{figure}

First, we have verified Eq. \eqref{inequality} starting from both an equilibrium and a nonequilibrium states as in Figs. \ref{fig1} and \ref{fig2}. 
For the case starting from an equilibrium state, we use $800,000$ independent samples, 
while we use $24,000$ samples for the case from a nonequilibrium state.
For the nonequilibrium setup, we use $t_0=1.0 \sqrt{m/k}$.  
These figures clearly exhibit that $\langle \Omega(t) \rangle_{\rm NE}$ which takes almost the identical values in both two initial conditions increases with time.
% regardless of the choice of an initial condition. 
From the absolute positivity of  $\langle \Omega(t) \rangle_{\rm NE}$, 
 it can play a role of the entropy production rate in nonequilibrium dissipative systems.

\begin{figure}
  \includegraphics[height=15em]{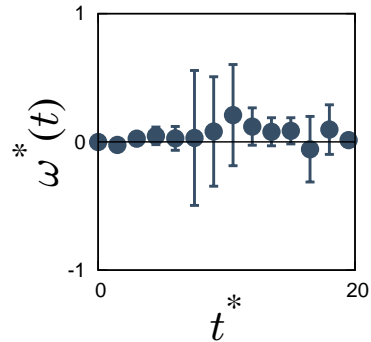}
  \caption{(Color online) The numerical verification of the integral fluctuation theorem Eq. \eqref{omega(t)} for $t_0 = 0.0$,
 where  $\omega^*(t) = \omega(t) \sqrt{m/k}$ is plotted against the scaled time $t* =  t \sqrt{k/m}$.}
  \label{fig3}
\end{figure}

\begin{figure}
  \includegraphics[height=15em]{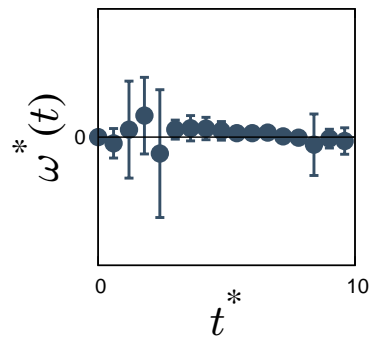}
  \caption{(Color online) The numerical verification of the integral fluctuation theorem Eq. \eqref{omega(t)} for $t_0 = 1.0\sqrt{m/k}$,
 where  $\omega^*(t) = \omega(t) \sqrt{m/k}$
  is plotted against the scaled time $t* =  t \sqrt{k/m}$.}
  \label{fig4}
\end{figure}

Figures \ref{fig3} and \ref{fig4} demonstrate that $\omega(t)$ introduced in Eq.(\ref{omega(t)}) keeps to be zero within the numerical accuracy regardless of the choice of an initial condition,
 where Fig. \ref{fig3} begins with an equilibrium condition ($t_0=0$), 
and Fig. \ref{fig4} starts from a nonequilibrium condition  ($t_0=1.0 \sqrt{m/k})$. 
 This result supports the validity of IFT \eqref{IFT} as well as its time differentiation Eq. \eqref{omega(t)}.  
 The results presented in Figs. 1-4 are remarkable, because,  as stated previously, the higher order correlations produces $\omega(t)=0$, 
 while  $\omega(t) \simeq \left <\Omega (t)\right >$ is expected to be held under the decoupling approximation.

We note that IFT (\ref{IFT}) contains all order of cumulants which is not suitable for numerical calculation.
Nevertheless, such an identity is useful to test the validity of an approximate theory such as perturbation expansion. 

\subsection{The verification of the generalized Green-Kubo formula}

Let us verify the generalized Green-Kubo formula (\ref{Green-Kubo2}) in this subsection.
Because we restrict our interest to the combination of steady processes as in Eq. (\ref{caseC}), the generalized Green-Kubo formula 
(\ref{Green-Kubo2}) should be held.
To verify the Green-Kubo formula we measure the average of the microscopic stress $\sigma_{xy}(t)$ in Eq. ({\ref{micro_stress}) . 
Figure \ref{fig5} shows the results of two cases with 
$t_0=0$ and $t_0=5.0 \sqrt{m/k}$,
where we use $N=800,000$ independent samples.
Although we have large fluctuations for the data with 
$t_0\ne 0$, 
we can conclude that the generalized Green-Kubo formula \eqref{Green-Kubo2} 
is still valid even if we start from a nonequilibrium initial condition. 
\begin{figure}
  \includegraphics[height=15em]{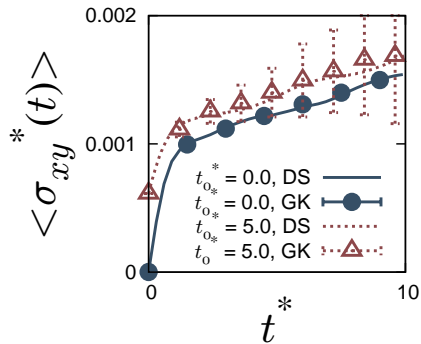}
  \caption{(Color online) The numerical verification of the generalized Green-Kubo formula in Eq.(\ref{Green-Kubo2}) for $t_0^* = t_0 \sqrt{k/m} = 0.0$ and $5.0$
  through the plot of   $\left < \sigma_{xy}^*(t) \right > = \left < \sigma_{xy}(t) \right > /(k/d)$ against the scaled time $t* =  t \sqrt{k/m}$, 
  where DS and GK, respectively, denote the numerical result in terms of the direct simulation of granular particles and the generalized Green-Kubo formula (\ref{Green-Kubo2}). 
  }
  \label{fig5}
\end{figure}

\section{Discussion}

In this paper, we obtain some exact nonequilibrium relations, though the generalized Green-Kubo formula is only held for steady dynamics.
To verify the validity, we have performed DEM of granular particles.
We should stress that the identities presented in this paper can be used above the jamming transition.
%The detailed numerical check on the validity of nonequilibrium identities will be reported elsewhere.\cite{otsuki}

Even after we confirm the validity of the nonequilibrium relations such as IFT and the generalized Green-Kubo formula, 
it is not easy to calculate the correlation function Eq. (\ref{Green-Kubo2}) analytically.
One of possible methods is to use the mode-coupling theory (MCT).
It is helpful to apply MCT for sheared liquids to characterize rheology near the jamming transition \cite{Suzuki-Hayakawa,suzuki13,HO2008,miyazaki02,fuchs02,miyazaki04,fuchs05,fuchs09}.
It is notable that Ref. \cite{Suzuki-Hayakawa} 
develops a linear response theory for a sheared thermostat system around a nonequilibrium steady state.
The application of this type of the response theory for granular fluids will be discussed elsewhere.

A different approach is to analyze the eigenvalue problem of the Liouville equation with the aid of full counting statistics \cite{sagawa2011}.
This is also a promising approach, but the eigenvalue problem of the Liouville equation for granular liquids is not easy.
We will also look for the possibility of this approach elsewhere.

We believe that Eq. (\ref{inequality}) is the most important result in this paper,
because this suggests that we may introduce the entropy-like quantity $S(\bs{\Gamma}(t))$ in Eq. (\ref{entropy}) for an arbitrary dissipative system.
%As stated in Sec.III, we have the relation $\rho(\bs{\Gamma},-t)=\exp[-\int_0\td\tau \Omega(t)]\rho_{\rm NE}(\bs{\Gamma})$ in general situations.
In the standard statistical mechanics, the differentiation of the entropy with respect to the energy gives the inverse temperature.
This idea still can be used for the statistical mechanics of  granular gases.

For sheared and jammed granular systems under a constant volume condition,  there are some papers to construct a statistical mechanics by using the stress ensemble \cite{henkes09,blumenfeld12,daniels13,max13} which might be the counter part of known Edwards ensemble \cite{Edwards89} for a system with a changing volume.
Indeed, there are some advantages to use the stress $\Sigma_{xy}(\bs{\Gamma}(t))=V\sigma_{xy}(\bs{\Gamma}(t))$ for sheared and jammed granular systems, 
because the energy itself is small, and the stress should be spatially uniform in a steady state, while the energy is localized.
Furthermore, we should indicate ${\rm Tr}\Sigma_{ij}$ is not far from the energy itself.  
Therefore, this paper may give some justification to introduce the effective temperature $T_{\rm eff}=\lim_{t\to \infty}(\partial S(\bs{\Gamma}(t))/\partial (\Sigma_{xy} (\bs{\Gamma}(t))\rangle_{\rm NE})^{-1}$ in the stress ensemble.
The possibility to construct a statistical mechanics of granular systems which can cover granular gases, stress ensembles and the Edwards ensemble  along this line will be discussed somewhere.

\section{Summary}

We obtained some exact relations  for dissipative classical particles.
We derived the integral fluctuation theorem (\ref{IFT}) and its equivalent expression (\ref{omega(t)}) around a nonequilibrium state, and confirmed the positivity of $\langle \Omega(\bs{\Gamma}(t)) \rangle_{\rm NE}$ as in Eq. (\ref{inequality}).
From IFT we confirmed the existence of an entropy-like quantity (\ref{entropy}) in an arbitrary dissipative system.
We also derived the conventional fluctuation theorem (\ref{conventional_FT}).
We gave a simple derivation of the Jarzynski equality (\ref{jarzynski}). % if the Hamiltonian depends on time explicitly.
We also obtained the generalized Green-Kubo formula (\ref{Green-Kubo2}) for the steady dynamics around a nonequilibrium state.
We numerically verified the validity of the obtained identities (\ref{omega(t)}) and (\ref{Green-Kubo2})  as well as Eq. (\ref{inequality}) for the case of sheared granular fluids.

\label{sec:summary}

\begin{acknowledgments}
The authors thanks S.-H. Chong, K. Suzuki, K. Saitoh, and S. Yukawa  for fruitful discussions. 
The authors also wish to thank Aspen Center for Physics, where parts of this work is developed.
This work was supported by 
the Grant-in-Aid of MEXT (Grant Nos. 25287098 and 25800220)
and in part by the Yukawa International Program for 
Quark-Hadron Sciences (YIPQS).
%The numerical computations were carried out in computers of YITP, Kyoto University.
\end{acknowledgments}

\appendix

%\section{Kawasaki idenity and related relations}

\section{Alternative derivation of the integral fluctuation theorem}

It is possible to obtain IFT from a different manner.
Substituting Eq. (\ref{kawasaki}) into Eq. (\ref{claasical-13-b}) under the initial condition (\ref{large-deviation}) for $t>0$ we obtain
\begin{eqnarray}\label{classical-ft3}
&&\langle \check{A}(\bs{\Gamma}(-t)) \rangle_{\rm NE}
=
\int d \Vect{\Gamma} A(\Vect{\Gamma}) \tilde{U}_\rightarrow(0,t) 
\rho_{\rm NE} (\Vect{\Gamma})  
\nonumber\\
&\quad& =
\int d \Vect{\Gamma} A(\Vect{\Gamma}) \left[e^{\int_0^td\tau \Lambda(\bs{\Gamma}(\tau))} \rho_{\rm NE}(\Vect{\Gamma}(t)) \right] .
\end{eqnarray}
%where we have used a simplified notation $\Lambda(\tau)\equiv \Lambda(\Vect{\Gamma}(t),t)$ with $\tau=t-0$ and used Eq.(\ref{2.2}).
Setting $\check{A}(\bs{\Gamma})=1$, the left hand side of Eq. (\ref{classical-ft3}) becomes one because of Eq. (\ref{claasical-13-b}).
On the other hand, if we adopt Eq. (\ref{large-deviation})
%the canonical distribution as the initial condition  
%$\rho_{\rm NE}(\Vect{\Gamma})=e^{- I(\mathbf{\Gamma})}/{\cal Z}$, 
with the aid of 
\begin{equation}\label{rho_time2}
\rho_{\rm NE}(\Vect{\Gamma}(t))= e^{-I(\bs{\Gamma}(t))}/{\cal Z}=%T_\rightarrow
 e^{- \int_{0}^tds \dot{{I}}(\mathbf{\Gamma}(s))}\rho_{\rm NE}(\Vect{\Gamma}) ,
\end{equation}
we obtain IFT (\ref{IFT}).

\section{Jarzynski equality of a gas starting from %the equilibrium condition
Eq.(\ref{eq:rho_eq})
}

In this Appendix, we only focus on the Jarzynski equality of a gas starting from an equilibrium initial condition (\ref{eq:rho_eq}) to illustrate the meaning of the generalized Jarzynski equality introduced in Sec.  III C. 
We also assume that the protocol parameter $\lambda(\tau)$ is proportional to the volume of system $V$. 
Here, we introduce the Hamiltonian $H_\lambda\equiv H(\bs{\Gamma}(\tau),\lambda(\tau))$ of the system for $0\le \tau \le t$.

Because the pressure $P$ is given by $P=-(\partial H_\lambda/\partial V)_S$ for an adiabatic process in thermodynamics,
the work $W(t)$ acting on the system during the volume change $V(\lambda(\tau))$ for $0\le \tau \le t$ can be represented by
$W(t)=\int_0^t d\tau \dot{V} (\partial H_\lambda/\partial V)_S=\int_0^t d\tau \dot{\lambda} (\partial H_\lambda/\partial \lambda)_S$,
where $\dot{X}=dX(\tau)/d\tau$ for an arbitrary $X(\tau)$.

Now, let us discuss the change of the energy in the system
\begin{eqnarray}
H(\bs{\Gamma}(t),1)-H(\bs{\Gamma}(0),0)
&=& \int_0^t d\tau \frac{d}{d\tau} H(\bs{\Gamma}(\tau),\lambda(\tau))
\nonumber\\
&=& \int_0^t d\tau \Huge\{ \dot{\lambda}(\tau) \frac{\partial H(\bs{\Gamma}(\tau),\lambda(\tau))}{\partial \lambda(\tau)} 
\nonumber\\
&& + \dot{\bs{\Gamma}}(\tau)\cdot \frac{\partial H(\bs{\Gamma}(\tau),\lambda(\tau))}{\partial \bs{\Gamma}(\tau)} \Huge\}
\nonumber\\
&=&
W(t)+Q(t),
\end{eqnarray}
where we have introduced the absorbing heat $Q(t)\equiv \int_0^t d\tau \dot{\bs{\Gamma}}(\tau)\cdot \frac{\partial H(\bs{\Gamma}(\tau),\lambda(\tau))}{\partial \bs{\Gamma}(\tau)}=
\int_0^t d\tau i \cl(\bs{\Gamma}(\tau)) H_\lambda $ of the system.
Note that $\dot{Q}(t)$ is given by
$\dot{Q}(t)=-\beta V \sigma_{xy}(\bs{\Gamma}(t))-2\beta R(\bs{\Gamma}(t))$ for sheared granular systems introduced in Sec. IV.

Let us introduce the partition function $Z(\beta,t)\equiv \int d\bs{\Gamma}(t) e^{-\beta H(\bs{\Gamma}(t),\lambda(t)=1)}$ with the initial inverse temperature $\beta$. 
The change of $Z(\beta,t)$  is associated with the change of the free energy $\Delta F$ as
\begin{eqnarray}\label{J-eq:from_H}
e^{-\beta \Delta F}&=& 
\frac{Z(\beta,t)}{Z(\beta,0)}=\frac{1}{Z(\beta,0)}\int d\bs{\Gamma}(t) e^{-\beta H(\bs{\Gamma}(t),1)}
\nonumber\\
&=&\int d\bs{\Gamma}
\left| \frac{\partial \bs{\Gamma}(t)}{\partial \bs{\Gamma}}\right|
\frac{e^{-\beta H(\bs{\Gamma},0)}}{Z(\beta,0)}
e^{-\beta \int_0^t d\tau \frac{d}{d\tau}H(\bs{\Gamma}(\tau),\lambda(\tau))}
\nonumber\\
&=& \left\langle  \exp\left[ -\int_0^t d\tau \Omega_{\rm eq}(\bs{\Gamma}(\tau),\lambda(\tau)) \right] \right\rangle_{\rm eq} ,
\end{eqnarray}
where $\Omega_{\rm eq}(\bs{\Gamma}(\tau),\lambda(\tau))=\beta \dot{W}(\tau)+\beta \dot{Q}(\tau)-\Lambda(\bs{\Gamma}(\tau))$.
If the system dynamics is non dissipative, we have the relation $\dot{Q}(\tau)=\Lambda(\bs{\Gamma}(\tau))=0$, then the relation 
(\ref{J-eq:from_H}) is reduced to the conventional Jarzynski equality.
For dissipative systems, it is reasonable that there exists a contribution from the effective absorbing heat $Q_{\rm eff}(t)=Q(t)-\beta^{-1}\int_0^t d\tau \Lambda(\bs{\Gamma}(\tau))=\beta^{-1}\int_0^t d\tau \Omega_{\rm eq}(\bs{\Gamma}(\tau),\lambda(\tau))$. In other words, the phase volume contraction is associated with the effective heat. 
This result is also reasonable, because the entropy production is given by $\int_0^t d\tau \Omega_{\rm eq}(\bs{\Gamma}(\tau),\lambda(\tau))$.

\section{Effect of measurement}

Recently, Sagawa and Ueda \cite{sagawa} have extended nonequilibrium identities to those under the effect of measurement.
The effect of measurement appears through the mutual information
\begin{equation}\label{mutual}
{\cal I}[\Vect{\Gamma}(t);y(t)]\equiv \ln \frac{P(y(t)|\Vect{\Gamma}(t))}{P(y(t))} ,
\end{equation}
where $y(t)$ represents a measurement outcome, %  $\Vect{\Gamma}_t$ is the abbreviation of $\Vect{\Gamma}(t)$
and $P(y(t)|\Vect{\Gamma}(t))$ is the conditional probability.
%This series of research is important because the information is coupled with thermodynamic quantities.
The original formulation is applied to Hamilton dynamics, but it is easy to apply to the dissipative dynamics.
 
The generalized Jarzynski equality for sheared granular liquids is given by
\begin{equation}\label{g-Jarzynski}
\left\langle \left\langle \exp\left[-\int_{0}^tds {\Omega}(\mathbf{\Gamma}(s))\right] %
e^{-{\cal I}(\bs{\Gamma})} e^{ \Delta {\cal F}} 
\right\rangle \right\rangle=1 ,
\end{equation}
where $\langle \langle A \rangle \rangle$ represents $\int d\Vect{\Gamma}\prod_t dy(t) \frac{e^{- {\cal I}(\mathbf{\Gamma})}}{{\cal Z}(0)}P(y(t)|\Vect{\Gamma}(t)) A $ for any function $A$.
Indeed, the left hand side of (\ref{g-Jarzynski}) is rewritten as
\begin{eqnarray}
{\rm LHS}&=& \int d\Vect{\Gamma}\prod_s dy(s) \frac{e^{- I(\mathbf{\Gamma},0)}}{{\cal Z}(0)}
P(y(s)|\Vect{\Gamma}(s))  \nonumber\\
&& \times
 \exp\left[- \int_{0}^tds\{\dot{I}(\mathbf{\Gamma}(s),\lambda(s))-
\Lambda(\mathbf{\Gamma}(s))\} \right] 
\nonumber\\
&& 
\times  \frac{P(y(s))}{P(y(s)|\Vect{\Gamma}(s))}\frac{{\cal Z}(0)}{{\cal Z}(t)} ,
\end{eqnarray}
where ${\cal Z}(t)=\int d\bs{\Gamma}e^{-I(\bs{\Gamma}(t),\lambda(t)=1)}$.
This is reduced to 1 by using $\int d\Vect{\Gamma}(t)=\int d\Vect{\Gamma} e^{\int_0^td\tau \Lambda(\mathbf{\Gamma}(\tau))} $.


\begin{thebibliography}{99}
%%%%%%%%%%%%%%%%%%%%%%%%%%%%%%%%%%%%%%%%%%%%%%%%%%%%%%%%%%%%%
% Some macros are available for the bibliography:
%  o for general use
%    \JL : general journals                 \andvol : Vol (Year) Page
%  o for individual journal 
%    \AJ   : Astrophys. J.           \NC         : Nuovo Cim.
%    \ANN  : Ann. of Phys.           \NPA, \NPB  : Nucl. Phys. [A,B]
%    \CMP  : Commun. Math. Phys.     \PLA, \PLB  : Phys. Lett. [A,B]
%    \IJMP : Int. J. Mod. Phys.      \PRA - \PRE : Phys. Rev. [A-E]     
%    \JHEP : J. High Energy Phys.    \PRL        : Phys. Rev. Lett.
%    \JMP  : J. Math. Phys.          \PRP        : Phys. Rep.
%    \JP   : J. of Phys.             \PTP        : Prog. Theor. Phys.     
%    \JPSJ : J. Phys. Soc. Jpn.      \PTPS       : Prog. Theor. Phys. Suppl.
% Usage:
%  \PRD{45,1990,345}          ==> Phys.~Rev.\ \textbf{D45} (1990), 345
%  \JL{Nature,418,2002,123}   ==> Nature \textbf{418} (2002), 123
%  \andvol{B123,1995,1020}    ==> \textbf{B123} (1995), 1020
%%%%%%%%%%%%%%%%%%%%%%%%%%%%%%%%%%%%%%%%%%%%%%%%%%%%%%%%%%%%%
  
%\bibitem{}

\bibitem{Zubarev74}
D.~N.~Zubarev, {\em Nonequilibrium Statistical Thermodynamics} (Consultants
  Bureau, New York, 1974).

\bibitem{McLennan88}
J.~A.~McLennan, {\em Introduction to Nonequilibrium Statistical Mechanics}
  (Prentice Hall, New Jersey, 1988).

%\bibitem{Sasa06}
%S.~Sasa and H.~Tasaki, \JL{J.~Stat.~Phys.,125,2006,125}

\bibitem{Evans08}
D.~J. Evans and G.~P. Morriss, {\em Statistical Mechanics of Nonequilibrium
  Liquids}, 2nd ed. (Cambridge University Press, Cambridge, 2008).

\bibitem{Morriss87}
G.~P. Morriss and D.~J. Evans, Phys. Rev. A \textbf{35}, 792 (1987).

\bibitem{FT93}
D.~J.~Evans, E.~G.~D.~Cohen, and G.~P.~Morriss, Phys. Rev. Lett. \textbf{71}, 2401 (1993).
\bibitem{GC95}
G.~Gallavotti and E.~G.~D.~Cohen, Phys. Rev. Lett. \textbf{74}, 2694 (1995).
\bibitem{Jarzynski}
C.~Jarzynski, Phys. Rev. Lett. \textbf{78}, 2690 (1997).
\bibitem{Kurchan}
J.~Kurchan, J.~Phys.~A: Math.~Gen., \textbf{31}, 3719 (1998).
\bibitem{Crooks}
G.~E.~Crooks, Phys. Rev. E \textbf{61}, 2361 (2000).
\bibitem{Hatano-Sasa}
T.~Hatano and S. I.~Sasa, Phys. Rev. Lett. {\bf 86}, 3463 (2001). 

\bibitem{Evans02}
D.~J. Evans and D.~J. Searles, Adv.~Phys., \textbf{51}, 1529 (2002).

\bibitem{Seifert12}
U. Seifert, Rep. Prog. Phys. {bf 75}, 126001 (2012).

\bibitem{Esposito09}
M. Esposito, U. Harbola, and S. Mukamel, Rev. Mod. Phys. {\bf 81}, 1665 (2009).

\bibitem{Ren10}
J. Ren, P. H\"{a}nggi, and B. Li, Phys. Rev. Lett. {\bf 104}, 170601 (2010).

\bibitem{kanazawa13} K. Kanazawa, T. Sagawa, and H. Hayakawa, Phys. Rev. E {\bf 87}, 052124 (2013).


\bibitem{mennon04} K. Feitosa and N. Menon, Phys. Rev. Lett. {\bf 92}, 164301 (2004).

\bibitem{kumar11} N. Kumar, S. Ramaswamy, and A. K. Sood, Phys. Rev. Lett. {\bf 106}, 118001 (2011).

\bibitem{joubaud12} S. Joubaud, D. Lohse, and D. van der Meer, Phys. Rev. Lett. {\bf 108}, 210604 (2012).

\bibitem{Naert12} A. Naert, EPL {\bf 97}, 20010 (2012).

\bibitem{mounier12} A. Mounier and A. Naert, EPL {\bf 100}, 30002 (2012).

\bibitem{Puglisi05} A. Puglisi, P. Visco, A. Barrat, E. Trizac, and F. van Wijland, Phys. Rev. Lett. {\bf 95}, 110202 (2005).

\bibitem{Puglisi05EPL}  A. Puglisi, P. Visco, E. Trizac, and F. van Wijland, EPL {\bf 72}, 55 (2005).

\bibitem{Puglisi06PRE} A. Puglisi, P. Visco, E. Trizac, and F. van Wijland, Phys. Rev. E {\bf 73}, 021301 (2006).

\bibitem{Puglisi06JSM} A. Puglisi, L. Rondoni, and A. Vulpiani, J. Stat. Mech. (2006) P08001.

\bibitem{Sarracino10} A. Sarracino, D. Villamaina, G. Gradenigo, and A. Puglisi, EPL {\bf 92}, 34001 (2010).

\bibitem{Evans94} D. J. Evans and D. J. Searles, Phys. Rev. E {\bf 50}, 1645 (1994).

\bibitem{Seifert05}
 U.~Seifert, Phys. Rev. Lett. \textbf{95}, 040602 (2005).



\bibitem{Chong09b}
S.-H.~Chong, M.~Otsuki, and H.~Hayakawa, Phys. Rev. E \textbf{81}, 041130 (2010).

\bibitem{Chong10}
S.-H.~Chong, M.~Otsuki, and H.~Hayakawa, 
Prog. Theor. Phys. Suppl. No.{\bf 184}, 77 (2010).

\bibitem{Hayakawa10a}
H. Hayakawa, S.-H. Chong, and M. Otsuki,
in IUTAM-ISIMM Symposium on Mathematical Modeling and Physical Instances of Granular Flow, pp.19-30
edited by J. D. Goddard, J. T. Jenkins and P. Govine (AIP vol.1227, New York, 2010).

%\bibitem{Hayakawa13} H. Hayakawa, arXiv:1303.2311.

%%ibitem{Komatsu08}
%T.~S.~Komatsu and N.~Nakagawa, Phys. Rev. Lett. \textbf{100}, 030601 (2008).

%\bibitem{Komatsu09}
%T.~S.~Komatsu, N.~Nakagawa, S.~Sasa and H.~Tasaki, J.~Stat.~Phys., \textbf{134}, 401 (2009).



%\bibitem{Jarzynski}
%C. Jarzynski, Phys. Rev. Lett. \textbf{78}, 2690 (1997).


\bibitem{brey97} J. J. Brey, J. W. Dufty, and A. Santos, J. Stat. Phys. {\bf 87} (1997), 1051.
\bibitem{dufty06} J. W. Dufty, A. Baskaran, and J. J. Brey, Phys. Rev. E {\bf 77}, 031310 (2008).
\bibitem{dufty06b} A. Baskaran, J. W. Dufty ,and J. J. Brey, Phys. Rev. E {\bf 77}, 031311 (2008).
\bibitem{HO2008} H. Hayakawa and M. Otsuki, Prog. Theor. Phys. {\bf 119}, 381 (2008).
\bibitem{Chong_etal_in_preparation} S.-H. Chong, K. Suzuki, M. Otsuki, and H. Hayakawa, in preparation.

\bibitem{Suzuki-Hayakawa} K. Suzuki and H. Hayakawa, Phys. Rev. E {\bf 87}, 012304 (2013).



\bibitem{vanNojie} T. P. C. van Noije, M. H. Ernst, and R. Brito, Physica A {\bf 251}, 266 (1998).



\bibitem{kawasaki} K. Kawasaki and J. D. Gunton, Phys. Rev. A {\bf 8}, 2048 (1973).


\bibitem{Evans90} D. J. Evans and G.~P. Morriss, {\em Statistical Mechanics of Nonequilibrium
  Liquids}, 1st ed. (Academic Press, London, 1990).




\bibitem{miyazaki02} K. Miyazaki and D. R. Reichman, Phys. Rev. E {\bf 66} (2002), 050501 (R).
\bibitem{fuchs02} M. Fuchs and M. E. Cates, Phys. Rev. Lett. {\bf 89}, 248304 (2002).
\bibitem{miyazaki04} K. Miyazaki, D. R. Reichman, and R. Yamamoto, Phys. Rev. E {\bf 70} (2004), 011501.
\bibitem{fuchs05} M. Fuchs and M. E. Cates, J. Phys.: Cond. Mat. {\bf 17} (2005), S1681.
\bibitem{fuchs09} M. Fuchs and M. E. Cates, J. Rheol. {\bf 53}, 957 (2009).

\bibitem{suzuki13} K. Suzuki and H. Hayakawa, AIP Conf. Proc. {\bf 1542}, 670 (2013).

%\bibitem{otsuki07} M. Otsuki and H. Hayakawa, J. Stat. Mech: Theor. Exp. (2009) P08003.
%\bibitem{brilliantov} N. V. Brilliantov and T. P\"{o}schel,  {\it Kinetic Theory of Granular Gases} (Oxford Univ. Press, Oxford, 2004).
%\bibitem{jenkins} J. T. Jenkins and M. W. Richman, Phys. Fluids {\bf 28} (1985) 3485.
%\bibitem{garzo} V. Garz\'{o} and J. W. Dufty, Phys. Rev. E {\bf 59} (1998) 5895.
%\bibitem{lutsko05} J. F. Lutsko, Phys. Rev. E {\bf 72} (2005) 021306.
%\bibitem{saitoh} K. Saitoh and H. Hayakawa, Phys. Rev. E {\bf 75}  (2007) 021302.
%\bibitem{hh-mo07} H. Hayakawa and M. Otsuki, Phys. Rev. E {\bf 76}  (2007) 051304.

%\bibitem{evans} D. J. Evans and G. P. Morriss, {\it Statistical Mechanics of Nonequilbrium Liquids} (Academic Press, London, 1990).



%\bibitem{resibois} P. M. V. Resibois and M. de Leener, Classical kinetic theory of fluids (John Wiley $\&$ Sons, New York, 1977).

%\bibitem{sagawa2011} T. Sagawa and H. Hayakawa, Phys. Rev. E {\bf 84}, 051110 (2011).
%
%\bibitem{saitoh2011}
%K. Saitoh and H. Hayakawa, Granular Matter \textbf{13}, 697 (2011).

%\bibitem{Sela}
%N. Sela, I. Goldhirsch and S. H. Noskowicz, Phys. Fluids {\bf 8}, 2337 (1996).




\bibitem{sagawa2011}
T. Sagawa and H. Hayakawa, Phys. Rev. E \textbf{84}, 051110 (2011).

\bibitem{henkes09} S. Henkes and B. Chakraborty, Phys. Rev. E {\bf 79}, 061301 (2009).
\bibitem{blumenfeld12} R. Blumenfeld, J. F. Jordan and S. F. Edwards, Phys. Rev. Lett. {\bf 109}, 238001 (2012).
\bibitem{daniels13} J. G. Puckett and K. E. Daniels, Phys. Rev. Lett. {\bf 110}, 058001 (2013).
\bibitem{max13} D. Bi, J. Zhang, R. P. Behringer and B. Chakraborty, EPL . {\bf 102}, 34002 (2013).

\bibitem{Edwards89} S. F. Edwards and R. A. Oakeshot, Physica A {\bf 157}, 1080- (1989).
%\bibitem{suzuki}

\bibitem{sagawa} T. Sagawa and M. Ueda, Phys. Rev. Lett. {\bf 104}, 090602 (2010).




\end{thebibliography}
\end{document}